\documentclass[11pt]{article}
\usepackage{amsmath,amssymb}
\begin{document}
\begin{flushright} SU-4252-785
\\
\end{flushright}         
\begin{center}
\vskip 3em
{\LARGE TOPOLOGY CHANGE FOR FUZZY PHYSICS: FUZZY 
\vskip 1em
SPACES AS HOPF ALGEBRAS}
\vskip 2em
{\large A. P. Balachandran$^\dagger$ and S. K\"{u}rk\c{c}\"{u}o\v{g}lu$^\dagger$
\footnote[1]{E-mails: bal@phy.syr.edu, skurkcuo@phy.syr.edu} 
\\[2em]}
\em{\oddsidemargin 0 mm
$^\dagger$Department of Physics, Syracuse University,
Syracuse, NY 13244-1130, USA}\\
\end{center}
\vskip 1em
\begin{abstract}
Fuzzy spaces are obtained by quantizing adjoint orbits of compact semi-simple Lie groups. Fuzzy spheres emerge from quantizing
$S^2$ and are associated with the group $SU(2)$ in this manner. They are useful for regularizing quantum field theories and 
modeling spacetimes by non-commutative manifolds. We show that fuzzy spaces are Hopf algebras and in fact have more structure
than the latter. They are thus candidates for quantum symmetries. Using their generalized Hopf algebraic structures, we can also 
model processes where one fuzzy space splits into several fuzzy spaces. For example we can discuss the quantum transition where
the fuzzy sphere for angular momentum $J$ splits into fuzzy spheres for angular momenta $K$ and $L$.
\end{abstract}

\newpage

\setcounter{footnote}{0}

\section{Introduction}

Fuzzy spaces provide finite-dimensional approximations to certain symplectic manifolds $M$ such as $S^2 \simeq {\mathbb C}P^1$, $S^2 
\times S^2$ and ${\mathbb C}P^2$. They are typically full matrix algebras \mbox{$Mat(N+1)$} of dimension \mbox{$(N+1) \times (N+1)$}.
The fuzzy sphere $S_F^2(J)$ for angular momentum $J= \frac{N}{2}$ for example is \mbox{$Mat(N+1)$}. As $N \rightarrow \infty$, a fuzzy
space provides an increasingly better approximation to the affiliated commutative algebra $C^\infty(M)$. Quantum field theories
(QFT's) on fuzzy spaces being finite, they are thus new regularizations of QFT's in the continuum.

Fuzzy spaces are obtained by quantizing adjoint orbits of compact Lie groups $G$. $S^2_F(J)$ is associated in this manner with $SU(2)$.

For a general algebra ${\mathcal A}$, it is not always possible to compose two of its representations $\rho$ and $\sigma$ to obtain a
third one. For groups such as $G$, we can do so and obtain the tensor product $\rho \otimes \sigma$. But for  ${\cal A}$, that is
possible only if  ${\cal A}$ is a coalgebra \cite{sweedler}. A coalgebra has a coproduct $\Delta$ which is a homomorphism 
from  ${\mathcal A}$ to ${\cal A} \otimes {\cal A}$ and the composition of its representations $\rho$ and $\sigma$ is the map 
$(\rho \otimes \sigma)\Delta$. If  ${\cal A}$ is more refined and is a Hopf algebra, then it closely resembles a group, in fact
sufficiently so that ${\mathcal A}$ can be used as a ``quantum symmetry group'' \cite{Mach-Schomerus}.

The group algebra $G^*$ consists of elements $\int_G d \mu(g) \alpha(g) g$ where $\alpha(g)$ is a smooth complex function
and $d \mu(g)$ is the $G$-invariant measure. It is isomorphic to the convolution algebra of functions on $G$.[see sec.3]. 
$G^*$ is a Hopf algebra.

In this paper, we establish that fuzzy spaces are irreducible representations $\rho$ of $G^*$ and inherit its Hopf algebra
structure. For fixed $G$, their direct sum is homomorphic to $G^*$. For example both $S_F^2(J)$ and \mbox{$\oplus_J S_F^2(J)
\simeq SU(2)^*$} are Hopf algebras. This means that we can define a coproduct on $S_F^2(J)$ and \mbox{$\oplus_J S_F^2(J)$}
and compose two fuzzy spheres preserving algebraic properties intact.

A group algebra $G^*$ and a fuzzy space from a group $G$ carry several actions of $G$. $G$ acts on $G$ and $G^*$
by left and right multiplications and by conjugation. Also for example, the fuzzy space $S_F^2(J)$ consists of $(2J+1) \times (2J+1)$
matrices and the spin $J$ representation of $SU(2)$ acts on these matrices by left and right multiplication and by conjugation.
The map $\rho$ of $G^*$ to a fuzzy space and the coproduct $\Delta$ are compatible with all these actions: they are
G-equivariant.

Elements $m$ of fuzzy spaces being matrices, we can take their hermitian conjugates. They are $*$-algebras if $*$ is hermitian
conjugation. $G^*$ also is a $*$-algebra. $\rho$ and $\Delta$ are $*$-homomorphisms as well: $\rho(\alpha^*)= 
\rho(\alpha)^\dagger$, $\Delta(m^*) = \Delta(m)^*$.

The last two properties of $\Delta$ on fuzzy spaces also derive from the same properties of $\Delta$ for $G^*$.

All this means that fuzzy spaces can be used as symmetry algebras. In that context however, $G$-invariance implies $G^*$-
invariance and we can substitute the familiar group invariance for fuzzy space invariance.

The remarkable significance of the Hopf structure seems to lie elsewhere. Fuzzy spaces approximate space-time algebras. $S_F^2(J)$ is 
an approximation to the Euclidean version of (causal) de Sitter space homeomorphic to $S^1 \times {\mathbb R}$, 
or for large radii of $S^1$, of Minkowski space \cite{Figari}. 
The Hopf structure then gives orderly rules for splitting and joining fuzzy spaces. The decomposition of $(\rho \otimes \sigma) 
\Delta$ into irreducible $*$-representations (IRR's) $\tau$ give fusion rules for states in $\rho$ and $\sigma$ combining to become 
$\tau$, while $\Delta$ on an IRR such as $\tau$ gives amplitudes for $\tau$ becoming $\rho$ and $\sigma$. In other words, $\Delta$ 
gives Clebsch-Gordan coefficients for space-times joining and splitting.
Equivariance means that these processes occur compatibly with $G$-invariance: $G$ gives selection rules for these processes in the
ordinary sense. The Hopf structure has a further remarkable consequence: An observable on a state in $\tau$ can be split into
observables on its decay products in $\rho$ and $\sigma$. 

There are similar results for field theories on $\tau$, $\rho$ and $\sigma$, indicating the possibility of many orderly calculations.

These mathematical results are very suggestive, but their physical consequences are yet to be explored.
  
The coproduct $\Delta$ on the matrix algebra Mat$(N+1)$ is not unique. Its choice depends on the group actions we care to preserve,
that of $SU(2)$ for $S_F^2$, $SU(N+1)$ for the fuzzy ${\mathbb C}P^N$ algebra 
${\mathbb C}P_F^N$ and so forth. It is thus the particular equivariance that determines the choice of $\Delta$.

We focus attention on the fuzzy sphere for specificity in what follows, but one can see that the arguments are valid for any fuzzy
space. Proofs for the fuzzy sphere are thus often assumed to be valid for any fuzzy space without comment.

Fuzzy algebras such as ${\mathbb C}P^N_F$ can be further ``$q$-deformed'' into certain quantum group algebras relevant for the study 
of $D$-branes. This theory has been developed in detail by Pawelczyk and Steinacker \cite{steinacker}. 

\section{Basics}

The canonical angular momentum generators of $SU(2)$ are $J_i \,(i=1,2,3)$. The unitary irreducible representations (UIRR's)
of $SU(2)$ act for any half-integer or integer $J$ on Hilbert spaces ${\cal H}^J$ of dimension $2J+1$. They have orthonormal basis 
$|J, M \rangle$, with $J_3 |J, M \rangle = M |J, M \rangle$ and obeying conventional phase conventions. The unitary matrix 
$D^J(g)$ of $g \in SU(2)$ acting on ${\cal H}^J$ has matrix elements \mbox{$\langle J ,M| D^J(g) |J, N \rangle = D^J(g)_{MN}$} 
in this basis.

Let 
\begin{equation}
V= \int_{SU(2)} d \mu (g)
\label{eq:volume}
\end{equation}
be the volume of $SU(2)$ with respect to the Haar measure $d \mu$. It is then well-known that \cite{Bal-Trahern} 
\begin{eqnarray}
\int_{SU(2)} d \mu (g) D^J(g)_{ij} \, D^K(g)_{kl}^\dagger &=& \frac{V}{2J+1} \, \delta_{JK} \, \delta_{il} \, 
\delta_{jk}\,, \\
\frac{2J+1}{V} \sum_{J \,, i j} D^J_{ij}(g) \, {\bar {D}}^J_{ij}(g^\prime) &=& \delta_g(g^\prime) \,,
\label{eq:comp}
\end{eqnarray}
where bar stands for complex conjugation and $\delta_g$ is the $\delta$-function on $SU(2)$ supported at $g$:
\begin{equation}
\int_{SU(2)} d \mu (g^\prime) \,\delta_g(g^\prime) \alpha(g^\prime) = \alpha(g)
\label{eq:deltaf}
\end{equation}
for smooth functions $\alpha$ on $G$.

We have also the Clebsch-Gordan series
\begin{equation}
D^K_{\mu_1 m_1} \, D^L_{\mu_2 m_2} = \sum_J C(K\,, L \,, J \,; \,\mu_1 \,, \mu_2) \, C(K \,, L \,, J\,; m_1 \,, m_2) \, 
D^J_{\mu_1 + \mu_2 \,, m_1+m_2}
\label{eq:clebschg}
\end{equation}
where $C$'s are the Clebsch-Gordan coefficients.

\section{The Group Algebra and the Convolution Algebra}

The group algebra consists of the linear combinations
\begin{equation}
\int_{G} d \mu(g) \, \alpha(g) \,g \,, \quad \quad d \mu(g)= \mbox{Haar \, measure \, on} \, G
\label{eq:groupalg}
\end{equation}
of elements $g$ of $G$, $\alpha$ being any smooth ${\mathbb C}$-valued function on $G$. The algebra product is induced from the
group product:
\begin{equation}
\int_G d \mu(g) \, \alpha(g) \,g \int_G d \mu(g^\prime) \, \beta(g^\prime) \,g^\prime :=
\int_G d \mu(g) \int_G d \mu(g^\prime) \alpha(g) \, \beta(g^\prime) (g g^\prime) \,.
\label{eq:groupproduct}
\end{equation}

We will henceforth omit the symbol $G$ under integrals. The right hand side of (\ref{eq:groupproduct}) is
\begin{equation}
\int d\mu(s) \, (\alpha *_c \beta)(s) \, s
\label{eq:conv}
\end{equation}
where $*_c$ is the convolution product:
\begin{equation}
(\alpha *_c \beta)(s) = \int d \mu(g) \alpha(g) \, \beta(g^{-1} s) \,.
\label{eq:convproduct}
\end{equation}
The convolution algebra consists of smooth functions $\alpha$ on $G$ with $*_c$ as their product. Under the map
\begin{equation}
\int d \mu (g) \alpha(g) g \rightarrow \alpha \,,
\label{eq:map1}
\end{equation}
(\ref{eq:groupproduct}) goes over to  $\alpha *_c \beta$ so that the group algebra and convolution algebra are isomorphic.
We call either as $G^*$. 

Using invariance properties of $d \mu$, (\ref{eq:map1}) shows that under the action
\begin{equation}
\int d \mu(g) \, \alpha(g) \,g \rightarrow
h_1  \, \left( \int d \mu(g) \alpha(g) g  \, \right) h_2^{-1} = \int d \mu(g) \alpha(g) h_1 g h_2^{-1} \,, \quad \quad h_i \in G \,,  
\label{eq:invariance1}
\end{equation}
$\alpha \rightarrow \alpha^\prime$ where 
\begin{equation}
\alpha^\prime(g) =\alpha (h_1^{-1} g h_2). 
\end{equation}
Thus the map (\ref{eq:map1}) is compatible with left- and right- $G$ actions. 

The group algebra is a $*$-algebra \cite{sweedler}, the $*$-operation being 
\begin{equation}
\left [ \int d \mu(g) \, \alpha (g) \, g \right ]^* = \int d \mu (g) \, {\bar \alpha (g)} g^{-1} \,.
\label{eq:starop}
\end{equation}
The $*$-operation in $G^*$ is 
\begin{equation} 
* : \alpha \rightarrow \alpha^* \,, \quad \alpha^*(g) = {\bar \alpha}(g^{-1}) \,.
\label{eq:qstarop1}
\end{equation}
Under the map (\ref{eq:map1}),
\begin{equation}
\left [\int d \mu (g) \alpha(g) g \right ]^* \rightarrow \alpha^*
\label{eq:map2}
\end{equation}
since
\begin{equation}
d \mu (g) = i \, Tr(g^{-1} \, d g) \wedge g^{-1} \, d g \wedge g^{-1} \, d g  = - d \mu (g^{-1}) \,.
\label{eq:measure1}
\end{equation}
(The minus sign in (\ref{eq:measure1}) is compensated by flips in ``limits of integration'', thus $\int d \mu (g) =
\int d \mu (g^{-1}) = V$.) Hence the map (\ref{eq:map1}) is a $*$-morphism, that is, it preserves ``hermitian conjugation''.

\section{The $*$-Homomorphism $G^* \rightarrow S_F^2$}

As mentioned earlier, henceforth we identify the group and convolution algebras and denote either by $G^*$.
We specialize to $SU(2)$ for simplicity.
We work with group algebra and and group elements, but one may prefer the convolution algebra instead for reasons of rigor.
(The image of $g$ is the Dirac distribution $\delta_g$ and not a smooth function.)

The fuzzy sphere algebra is not unique, but depends on the angular momentum $J$ as shown by the 
notation $S_F^2(J)$, which is $Mat(2J+1)$. Let
\begin{equation}
{\cal S}_F^2 = \oplus_{J} S_F^2(J) = \oplus_J Mat(2J+1) \,.
\label{eq:fuzzys}
\end{equation}

Let $\rho (J)$ be the unitary irreducible representation of angular momentum $J$ for $SU(2)$:
\begin{equation}
\rho (J) : \quad g \rightarrow \langle \rho (J) , g \rangle := D^J(g) \,.
\label{eq:uirr1}
\end{equation}
We have
\begin{equation}
\langle \rho (J) , g \rangle \, \langle \rho (J) , h \rangle = \langle \rho (J) , g h \rangle \,.
\label{eq:uirr2}
\end{equation}
Choosing the $*$-operation on $D^J(g)$ as hermitian conjugation, $\rho (J)$ extends by linearity to a $*$-homomorphism 
on $G^*$:
\begin{gather}
\Big \langle \rho (J) , \int d \mu (g) \alpha (g) g \Big \rangle = \int d \mu (g) \alpha (g) D^J(g) \nonumber \\
\Big \langle \rho (J) , \Big(\int d \mu (g) \alpha (g) g \Big)^* \Big \rangle = \int d \mu (g) \bar {\alpha}(g) D^J(g)^\dagger \,. 
\label{eq:PCSU}
\end{gather}
$\rho (J)$ is also compatible with group actions on $G^*$ (that is, it is equivariant with respect to these actions):
\begin{equation}
\Big \langle \rho (J) , \int d \mu (g) \alpha (g) h_1 g h_2^{-1} \Big \rangle= \int d \mu (g) \alpha (g) D^J(h_1)  D^J(g)  
D^J(h_2^{-1}) \,, \quad h_i \in SU(2)  
\label{eq:equivariance1}
\end{equation}

As by (2),
\begin{gather}
\Big \langle \rho (J), \frac{2K+1}{V} \int d \mu (g) (D^K_{ij})^\dagger (g) g \Big \rangle = e^{j i} (J) \delta_{KJ} \,, 
\nonumber \\
e^{j i} (J)_{rs} = \delta_{jr} \delta_{is} \,, \quad i,j,r,s \in [-J, \cdots 0 \cdots, J] \,,
\label{eq:PCSU2}
\end{gather}
we see by (\ref{eq:uirr2}) and (\ref{eq:PCSU}) that $\rho(J)$ is a $*$-homomorphism from $G^*$
to \mbox{$S_F^2(J) \oplus \lbrace 0 \rbrace$}, where $\lbrace 0 \rbrace$ denotes the zero elements of $\oplus_{K \neq J} S_F^2(K)$,
the $*$-operation on $S_F^2(J)$ being hermitian conjugation. Identifying $S_F^2(J) \oplus \lbrace 0 \rbrace$ with $S_F^2(J)$, 
we thus get a $*$-homomorphism $ \rho (J) : G^* \rightarrow S_F^2(J)$. It is also seen to be equivariant with respect 
to $SU(2)$ actions, they are given on the basis $e^{ji}(J)$ by $D^J(h_1) e^{ji}(J) D^J(h_2)^{-1}$.

We can think of (\ref{eq:PCSU}) as giving a map
\begin{equation}
\rho : g \quad \rightarrow  \quad \langle \rho (.) \,, g \rangle := g(.)
\label{eq:Pmap} 
\end{equation}
to a matrix valued function $g(.)$ on the space of UIRR's of $SU(2)$ where
\begin{equation}
g(J) = \langle \rho (J) \,, g \rangle \,.
\label{eq:Pmap2}
\end{equation}
The homomorphism property (\ref{eq:PCSU}) is expressed as the product $g(.) h(.)$ of these functions where
\begin{equation}
g(.) h(.) (J) = g(J) h(J)
\label{eq:propoint}
\end{equation}
is the point-wise product of matrices. This point of view is helpful for later discussions.

As emphasized earlier, this discussion works for any group $G$, its UIRR's, and its fuzzy spaces barring technical problems. 
Thus $G^*$ is $*$-isomorphic to the $*$-algebra of functions $g(.)$ on the space of its UIRR's $\tau$, with $g(\tau) 
= D^{\tau}(g)$, the linear operator of $g$ in the UIRR $\tau$ and $g^*(\tau) = D^{\tau}(g)^\dagger$.

A fuzzy space is obtained by quantizing an adjoint orbit $G/H$, $H \subset G$ and approximates $G/H$. It is a full matrix algebra
associated with a particular UIRR $\tau$ of $G$. There is thus a $G$-equivariant $*$-homomorphism from $G^*$ to the fuzzy 
space.

At this point we encounter a difference with $S_F^2(J)$. For a given $G/H$ we generally get only a subset of UIRR's $\tau$. For
example ${\mathbb C}P^2 = SU(3)/U(2)$ is associated with just the symmetric products of just $3$'s (or just $3^*$'s) of $SU(3)$.
Thus the direct sum of matrix algebras from a given $G/H$ is only homomorphic to $G^*$. 

Henceforth we call the space of UIRR's of $G$ as ${\hat G}$. For a compact group, ${\hat G}$ can be identified with the set of 
discrete parameters specifying all UIRR's.

\section{$G^*$ is a Hopf Algebra}

$G^*$ has more significant structures. It is in particular a Hopf algebra. That means that it has a coproduct $\Delta$,
a counit $\varepsilon$ and an antipode $S$ (induced from the canonical Hopf algebra structure) defined as follows
\begin{eqnarray}
\Delta (g) &=& g \otimes g \,,
\label{eq:co1} 
\\
\varepsilon (g) &=& {\bf 1} \in {\mathbb C} \,, 
\label{eq:counit}
\\  
S(g) &=& g^{-1} \,.
\label{eq:hopfcop1}
\end{eqnarray}
Here $\varepsilon$ is the one-dimensional trivial representation of $G$ and $S$ maps $g$ to its inverse. $\Delta$, $\varepsilon$ and
$S$ fulfill all the axioms of a Hopf algebra as is easy to verify.

The properties of a group $G$ are captured by the algebra of matrix-valued functions $g(.)$ on ${\hat G}$ with point-wise 
multiplication, this algebra being isomorphic to $G^*$. In terms of $g(.)$, (\ref{eq:co1} - \ref{eq:hopfcop1}) translate to 
\begin{eqnarray}
\Delta \big(g(.)\big) &=& g(.) \otimes g(.) \,, \\
\varepsilon (g(.)) &=& {\bf 1} \in {\mathbb C} \,, \\  
S\big( g(.) \big) &=& g^{-1}(.) \,.
\label{eq:hopfcop2}
\end{eqnarray}   
Note that $ g(.) \otimes g(.)$ is a function on ${\hat G} \otimes {\hat G}$.

\section{Hopf Algebra for the Fuzzy Spaces}

Any fuzzy space has a Hopf algebra, we show it here for the fuzzy sphere.

Let $\delta_J$ be the $\delta$-function on ${\widehat {SU(2)}}$: 
\begin{equation}
\delta_J(K) := \delta_{JK} \,.
\end{equation}
(Since the sets of $J$ and $K$ are discrete we have Kronecker delta and not a delta function).

Then 
\begin{equation}
e^{ji}(J) \, \delta_J = \frac{2J+1}{V} \int d \mu (g) D_{ij}^J(g)^\dagger g(.)
\end{equation}
Hence 
\begin{equation}
\Delta ( e^{ji}(J) \delta_J) = \frac{2J+1}{V} \int d \mu (g) D_{ij}^J(g)^\dagger g(.) \otimes g(.) \,.
\label{eq:edeltaj}
\end{equation}
At $(K, L) \in {\widehat {SU(2)}} \otimes {\widehat {SU(2)}}$, this is
\begin{equation}
\Delta \big(e^{ji}(J) \big)(K,L) = \frac{2J+1}{V} \int d \mu (g) D_{ij}^J(g)^\dagger \, D^K(g) \otimes D^L(g) \,.
\label{eq:KL}
\end{equation}

As $\delta_J^2 = \delta_J$ and $\delta_J e^{ji}(J) = e^{ji}(J) \delta_J$, we can identify $e^{ji}(J) \delta_J$ with
$e^{ji}(J)$:
\begin{equation}
e^{ji}(J) \delta_J \simeq e^{ji}(J) \,.
\end{equation}
Then (\ref{eq:edeltaj}) or (\ref{eq:KL}) show that there are many coproducts $\Delta = \Delta_{KL}$ we can define and they are 
controlled by the choice of $K$ and $L$:
\begin{equation}
\Delta \big (e^{ji}(J) \delta_J \big)(K,L) := \Delta_{KL} \big( e^{ji}(J) \big).
\label{eq:copKL}
\end{equation}

Technically a coproduct $\Delta$ is a homomorphism from an algebra ${\cal A}$ to ${\cal A} \otimes {\cal A}$ so that
only $\Delta_{JJ}$ is a coproduct. But we will call all $\Delta_{KL}$ as coproducts.

It remains to simplify the RHS of (\ref{eq:KL}). Using (\ref{eq:clebschg}), (\ref{eq:KL}) can be written as
\begin{multline}
\Delta \big (e^{ji}(J) \delta_J \big)_{\mu_1 \mu_2 \,, m_1 m_2} = \frac{2J+1}{V} \int d \mu (g) D_{ij}^J(g)^\dagger \,  
\sum_{J^\prime} C(K,L,J^{\prime}; \mu_1 \,, \mu_2) \\ 
\times C(K, L, J^{\prime}; m_1 \,, m_2) \, D^{J^{\prime}}_{\mu_1 + \mu_2 \,, m_1 + m_2} \,,
\label{eq:copexp}
\end{multline}
with $\mu_1 \,, \mu_2$ and $m_1 \,, m_2$ being row and column indices. The RHS of (\ref{eq:copexp}) is
\begin{multline}
C(K,L,J; \mu_1 \,, \mu_2) \, C(K, L, J; m_1 \,, m_2) \delta_{j \,, \mu_1+\mu_2} \delta_{i, m_1+m_2} \\
= \sum_{\substack{\mu^{\prime}_1+\mu^{\prime}_2 = j \\ 
m^{\prime}_1+m^{\prime}_2= i}} C(K,L,J; \mu^{\prime}_1 \,, \mu^{\prime}_2) \, C(K, L, J; m^{\prime}_1 \,, m^{\prime}_2) \,
\big(e^{\mu^{\prime}_1 m^{\prime}_1} (K) \big)_{\mu_1 m_1} \otimes \big(e^{\mu^{\prime}_2 m^{\prime}_2} (L) \big)_{\mu_2 m_2} \,.
\label{eq:copexp2}  
\end{multline}
Hence we have the coproduct
\begin{equation}
\Delta_{KL} \big( e^{ji}(J) \big) = \sum_{\substack{\mu_1+\mu_2 = j \\ m_1+m_2= i}}
C(K,L,J; \mu_1 \,, \mu_2) \, C(K, L, J; m_1 \,, m_2) \, e^{\mu_1 m_1}(K) \otimes e^{\mu_2 m_2}(L) \,.
\label{eq:copKL2}
\end{equation}

Writing $C(K,L,J; \mu_1 \,, \mu_2\,, j) =C(K,L,J; \mu_1 \,, \mu_2) \delta_{\mu_1+\mu_2 \,, j}$ for the first Clebsch-Gordan 
coefficient, we can delete the constraint $j=\mu_1+\mu_2$ in summation. $C(K,L,J; \mu_1 \,, \mu_2 \,, j)$ is an invariant tensor 
when $\mu_1 \,, \mu_2$ and $j$ are transformed appropriately by $SU(2)$. Hence (\ref{eq:copKL2}) is preserved by $SU(2)$ action
on $j, \mu_1, \mu_2$. The same is the case for $SU(2)$ action on $i, m_1, m_2$. In other words, the coproduct in (\ref{eq:copKL2}) 
is equivariant with respect to both $SU(2)$ actions. 

Since any $M \in Mat(2J+1)$ is $\sum_{i,j} M_{ji} e^{ji}(J)$, (\ref{eq:copKL2}) gives
\begin{equation}
\Delta_{KL}(M)= \sum_{\mu_1 \,,\mu_2\,, m_1 \,, m_2} C(K,L,J; \mu_1 \,, \mu_2) \, C(K, L, J; m_1 \,, m_2)
M_{\mu_1+\mu_2 \,, m_1+m_2} e^{\mu_1 m_1}(K) \otimes e^{\mu_2 m_2}(L) \,.
\label{eq:cobasic}
\end{equation}
This is the basic formula. It preserves conjugation $*$ (induced by hermitian conjugation of matrices):
\begin{equation}
\Delta(M^\dagger) = \Delta(M)^\dagger \,.
\label{eq:cobasic2}
\end{equation}

It can be directly checked from (\ref{eq:cobasic}) that $\Delta_{KL}$ is a homomorphism, that is that $\Delta_{KL}(MM^\prime)
= \Delta_{KL}(M) \Delta_{KL}(M^\prime)$. 

It remains to record the fuzzy analogues of counit $\varepsilon$ and antipode $S$. For the counit we have
\begin{eqnarray}
\varepsilon \big(e^{ji}(J) \delta_J\big) &=& \frac{2J+1}{V} \int d \mu (g) D_{ij}^J(g)^\dagger \varepsilon \big( g(.) \big) 
\nonumber \\
&=& \frac{2J+1}{V} \int d \mu (g) D_{ij}^J(g)^\dagger {\bf 1} \nonumber \\
&=& \frac{2J+1}{V} \int d \mu (g) D_{ij}^J(g)^\dagger D^0(g) \,.
\end{eqnarray} 
Using equation (2) and the fact that $D^0(g)$ is a unit matrix with only one entry which we denote by
$00$, we have 
\begin{equation}
\varepsilon \big(e^{ji}(J) \delta_J \big)_{00}(K) = \delta_{0J} \delta_{j0} \delta_{i0}  \,, \quad \forall K \in 
{\widehat {SU(2)}} \,.
\end{equation}

For the antipode, we have
\begin{eqnarray}
S \big(e^{ji}(J) \delta_J\big) &=& \frac{2J+1}{V} \int d \mu (g) D_{ij}^J(g)^\dagger S \big( g(.) \big) \nonumber \\
&=& \frac{2J+1}{V} \int d \mu (g) D_{ij}^J(g)^\dagger g^{-1}(.) 
\label{eq:antipode1}
\end{eqnarray}
or
\begin{equation}
S \big(e^{ji}(J) \delta_J\big)(K) = \frac{2J+1}{V} \int d \mu (g) D_{ij}^{J}(g)^\dagger D^K(g^{-1}) \,.
\label{eq:antipode2}
\end{equation} 

Now let $\vec{J}$ denote the angular momentum operator in the UIRR $K$ and $C=e^{-i \pi J_2}$ be the charge conjugation matrix. 
It fulfills $C D^K(g) C^{-1} = {\bar D}^K(g)$. Then since $D^K(g^{-1}) = D^K(g)^\dagger$,
\begin{equation}
D^K(g^{-1}) = C D^K(g)^T C^{-1} \,,
\end{equation}
where $T$ denotes transposition. We insert this in (\ref{eq:antipode2}) and use (2) to find  
\begin{eqnarray}
S \big(e^{ji}(J) \delta_J\big)_{k \ell}(K) &=& \frac{2J+1}{V} \int d \mu (g) D_{ij}^J(g)^\dagger 
\big ( C_{ku} D^K(g)^T_{u \upsilon} C^{-1}_{\upsilon \ell} \big) \nonumber \\
&=& \frac{2J+1}{V} \int d \mu (g) D_{ij}^J(g)^\dagger C_{ku} D^K(g)_{\upsilon u} C^{-1}_{\upsilon \ell} \nonumber \\ 
&=& \delta _{JK} C_{ku} \delta_{ui} \delta_{\upsilon j} C^{-1}_{\upsilon \ell} \nonumber \\
&=& \delta_{JK} C_{ki} C^{-1}_{j \ell} \,.
\end{eqnarray}
This can be simplified further. Since in the UIRR $K$,
\begin{equation}
\big(e^{- i \pi J_2}\big)_{ki} = \delta_{-ki}(-1)^{K+k} = \delta_{-ki}(-1)^{K-i}  \,,
\end{equation}
and $C^{-1}=C^T$, we find
\begin{eqnarray}
S \big(e^{ji}(J) \delta_J\big)_{k \ell}(K) &=& \delta_{JK} \delta_{-ki} \delta_{-\ell j} (-1)^{2K-i-j} \nonumber \\
&=& \delta_{JK} (-1)^{2J-i-j} e^{-i\,,-j}(J)_{k \ell} \,.
\end{eqnarray}
Thus
\begin{equation}
S \big(e^{ji}(J) \delta_J\big)(K) = \delta_{JK} (-1)^{2J-i-j} e^{-i\,,-j}(J) \,.
\end{equation}

\section{Interpretation}

The matrix $M$ can be interpreted as the wave function of a particle on the spatial slice $S_F^2(J)$. The scalar product on these
wave functions is given by $(M, N ) = Tr M^\dagger N$, $M, N \in S_F^2(J)$.

We can also regard $M$ as a fuzzy two-dimensional Euclidean scalar field or even as a field on a spatial slice $S_F^2(J)$ of a 
three dimensional space-time $S_F^2(J) \times {\mathbb R}$.

Let us look at the particle interpretation. Then (\ref{eq:cobasic}) gives the amplitude, up to an overall factor, for $M \in
S_F^2(J)$ splitting into a superposition of wave functions on $S_F^2(K) \otimes S_F^2(L)$. It models the process where a fuzzy 
sphere splits into two others \cite{BalPaulo}. The overall factor is the reduced matrix element much like the reduced 
matrix elements in angular momentum selection rules. It is unaffected by algebraic operations on    
$S_F^2(J), S_F^2(K)$ or $S_F^2(L)$ and is determined by dynamics. 

Now (\ref{eq:cobasic}) preserves trace and scalar product:
\begin{gather}
Tr \Delta_{KL}(M) = Tr M \,, \nonumber \\
\big( \Delta_{KL}(M), \Delta_{KL}(N) \big) = (M, N) \,. 
\label{eq:traces}
\end{gather}
So (\ref{eq:cobasic}) is a unitary branching process. This means that the overall factor is a phase.

$\Delta_{KL}(S_F^2(J))$ has all the properties of $S_F^2(J)$. So (\ref{eq:cobasic}) is also a precise rule on how $S_F^2$ sits in
$S_F^2(K) \otimes S_F^2(L)$. We can understand ``how $\Delta_{KL}(M)$ sits'' as follows. A basis for 
$S_F^2(K) \otimes S_F^2(L)$ is $e^{\mu_1 m_1} (K) \otimes e^{\mu_2 m_2} (L)$. We can choose another basis where 
left- and right- angular momenta are separately diagonal by 
coupling $\mu_1$ and  $\mu_2$ to give angular momentum $\sigma \in \lbrack 0, \frac{1}{2}, 1, \hdots , K+L \rbrack$, and $m_1$ 
and $m_2$ to give 
angular momentum $\tau \in \lbrack 0, \frac{1}{2}, 1, \hdots , K+L \rbrack$. In this basis, $\Delta_{KL}(M)$ is zero except 
in the block with $\sigma = \tau =J$.    

So the probability amplitude for $M \in S_F^2(J)$ splitting into $P \otimes Q \in S_F^2(K) \otimes S_F^2(L)$ for normalized wave
functions is 
\begin{equation}
phase \times Tr(P \otimes Q)^\dagger \Delta_{KL}(M) \,.
\label{eq:probamp1}
\end{equation}  
Branching rules for different choices of $M, P$ and $Q$ are independent of the constant phase and can be determined.

Written in full, (\ref{eq:probamp1}) is seen to be just the coupling conserving left- and right- angular momenta of 
$P^\dagger, Q^\dagger$ and $M$. That alone determines (\ref{eq:probamp1}).

An observable $A$ is a self-adjoint operator on a wave function $M \in S_F^2(J)$. Any linear operator on $S_F^2(J)$ can be written as
$\sum B_\alpha^L C_\alpha^R$ where $B_\alpha \,, C_\alpha \in S_F^2(J)$ and $B_\alpha^L$ and $C_\alpha^R$ act by left- and right-
multiplication: $B_\alpha^L M = B_\alpha M \,, C_\alpha^R M = M C_\alpha$. Any observable on $S_F^2(J)$ has an action on its branched
image $\Delta_{KL} (S_F^2(J))$:
\begin{equation}
\Delta_{KL} (A) \Delta_{KL}(M) :=\Delta_{KL} (AM) \,.    
\label{eq:obaction}
\end{equation}
By construction, (\ref{eq:obaction}) preserves algebraic properties of operators. $\Delta_{KL} (A)$ can actually act on all of 
$S^2_F(K) \otimes S_F^2(L)$, but in the basis described above it is zero on vectors with $\sigma \neq J$ and/or $\tau \neq J$.

This equation is helpful to address several physical questions. For example if $M$ is a wave function with a
definite eigenvalue for $A$, then $\Delta_{KL}(M)$ is a wave function with the same eigenvalue for $\Delta_{KL}(A)$. This follows from
$\Delta_{KL}(BM) = \Delta_{KL}(B) \Delta_{KL}(M)$ and $\Delta_{KL}(MB) = \Delta_{KL}(M) \Delta_{KL}(B)$. Combining this with 
(\ref{eq:traces}) and the other observations, we see that mean value of $\Delta_{KL}(A)$ in $\Delta_{KL}(M)$ and of $A$ in $M$ are
equal.

In summary all this means that every operator on $S_F^2(J)$ is a constant of motion for the branching process (\ref{eq:cobasic}).

Now suppose $R \in S_F^2(K) \otimes S_F^2(L)$ is a wave function which is not necessarily of the form $P \otimes Q$. 
Then we can also give a formula for the probability amplitude
for finding $R$ in the state described by $M$. Note that $R$ and $M$ live in different fuzzy spaces. The answer is 
\begin{equation}
constant \times Tr R^\dagger \Delta_{KL}(M) \,.
\end{equation}

If $M, P, Q$ are fields with $S_F^2(I) \, (I= J,K,L)$ a spatial slice or space-time, (\ref{eq:probamp1}) is an interaction of 
fields on different fuzzy manifolds. It can give dynamics to the branching process of fuzzy topologies discussed above.

\section{The Pre\v{s}najder Map}

This section is somewhat disconnected from the material in the rest of the paper.

$S_F^2(J)$ can be realized as an algebra generated by the spherical harmonics $Y_{lm} \, (l \leq 2J)$ which are functions on the 
two-sphere $S^2$. Their product is not point-wise. It can be the coherent state (or Voros \cite{voros}) $*_c$ or Moyal $*_M$ product.

But we saw that $S_F^2(J)$ is isomorphic to the convolution algebra of functions $D_{MN}^J$ on $SU(2) \simeq S^3$. 

It is reasonable to wonder how functions on $S^2$ and $S^3$ get related preserving the respective algebraic properties.

The map connecting these spaces is described by a function on $SU(2) \times S^2 \approx S^3 \times S^2$ and was first introduced 
by Pre\v{s}najder \cite{peter}. We give its definition and introduce its properties here. It generalizes to any group $G$.

Let $a_i \,, a_j^\dagger$ $(i = 1,2)$ be Schwinger oscillators for $SU(2)$ and let 
\begin{equation}
|z \rangle_J = \frac{(z_i a_i^\dagger)^{2J}}{\sqrt{2J!}} \,|0 \rangle \,, \quad \sum|z_i|^2 =1 \,
\label{eq:coherentstate1}
\end{equation}
be the normalized Perelomov vectors \cite{perelomov}. If $U(g)$ is the unitary operator implementing $g \in SU(2)$ in the 
spin $J$ UIRR, the Pre\v{s}najder function \cite{peter} $P_J$ is given by 
\begin{gather}
P_J(g, \vec{n})= \langle z|U(g)|z \rangle_J = D_{JJ}^J(h^{-1} g h)\,, \nonumber \\      
{\vec n} = z^\dagger {\vec \tau} z \,, \quad  {\vec n} \cdot {\vec n} =1 \,, \quad h = \left (
\begin{array}{cc}
z_1 & -{\bar z}_2 \\
z_2 & {\bar z}_1 \\
\end{array}
\right ) \,.
\label{eq:presnajdermap}
\end{gather}
Now ${\vec n} \in S^2$. As the phase change $z_i \rightarrow z_i e^{i \theta}$ does not effect $P_J$, besides $g$, it 
depends only on ${\vec n}$. It is a function on
\begin{equation}
\big( SU(2) \simeq S^3 \big) \times \big \lbrack SU(2)/U(1) \big \rbrack \simeq S^3 \times S^2 \,.
\label{eq:domain1}
\end{equation}

A basis of $SU(2)$ functions for spin $J$ is $D_{ij}^J$. A basis of $S^2$ functions for spin $J$ is $E^{ij}(J \,, .)$ where 
\begin{equation}
E^{ij} (J \,, {\vec n}) = \langle z| e^{ij}(J) |z \rangle_J = D^J(h^{-1})_{Ji} D^J(h)_{jJ} \,, \quad  
\mbox{no \, sum \, on} \, J \,.
\label{eq:basisonS2}
\end{equation}
The transform of $D_{ij}^J$ to $E^{ij}(J,.)$is given by
\begin{equation}
E^{ij}(J \,, {\vec n}) = \frac{(2J+1)}{V} \int d \mu(g) {\bar P_J}(g, {\vec n}) D_{ij}^J(g) \,.
\label{eq:transforms}
\end{equation}
This can be inverted by constructing a function $Q_J$ on $SU(2) \times S^2$ such that 
\begin{equation}
\int_{S^2} d \Omega ({\vec n}) Q_J(g^\prime \,, {\vec n}) {\bar P}_J (g \,, {\vec n})
= \sum_{ij} D_{ij}^J(g^\prime) {\bar D}_{ij}^J(g) \,, \quad \quad d \Omega ({\vec n}) 
= \frac{d \cos \theta d \varphi}{4 \pi} \,,
\label{eq:qfunction}
\end{equation}
$\theta$ and $\varphi$ being the polar and azimuthal angles on $S^2$. Then using (2), we get
\begin{equation}
D_{ij}^J (g^\prime) = \int_{S^2} d \Omega ({\vec n}) Q_J(g^\prime \,, {\vec n}) 
E^{ij}(J \,, {\vec n})\,.  
\label{eq:transform1}
\end{equation}

Consider first $J = \frac{1}{2}$. In that case 
\begin{equation}
{\bar P}_{\frac{1}{2}} (g \,, {\vec n}) = {\bar g}_{kl} {\bar z}_k z_l
={\bar g}_{kl} \Big( \frac{1 + {\vec \sigma} \cdot {\vec n}}{2} \Big)_{lk}
\label{eq:presnajdermap1/2}
\end{equation}
where $g$ is a $2 \times 2$ $SU(2)$ matrix and $\sigma_i$ are Pauli matrices. Since
\begin{equation}
\int_{S^2} d \Omega ({\vec n}) n_i n_j = \frac{1}{3} \delta_{ij} \,,   
\label{eq:intid1}
\end{equation}
we find 
\begin{eqnarray}
Q_{\frac{1}{2}} (g^\prime \,, {\vec n}) &=& Tr {\tilde g}^\prime (1+ 3  {\vec \sigma} \cdot {\vec n}) \,, \\
g^\prime &=& 2 \times 2 \, \mbox{SU(2) \, matrix} \,, \nonumber \\
{\tilde g}^\prime &=& \mbox{transpose \, of} \, g^\prime \,. \nonumber   
\label{eq:q1/2}
\end{eqnarray}

If $J=\frac{N}{2}$, $D^J(g)$ acts on the symmetric product on $N$ ${\mathbb C}^2$'s and can be written as 
$\underbrace{g \otimes g \otimes \cdots \otimes g}_{N \, factors}$ and (\ref{eq:presnajdermap1/2}) gets replaced by
\begin{equation}
{\bar P}_J (g \,, {\vec n}) = \Big \lbrack Tr {\bar g} \Big( \frac{1 + {\vec \sigma} 
\cdot {\vec n}}{2} \Big) \Big \rbrack^N \,.  
\end{equation}
Then $Q_J (g^\prime \,, {\vec n})$ is defined by (\ref{eq:qfunction}). It exists. We have not found a neat formula
for it. 

As the relation between $E^{ij}$ and $Y_{lm}$ can be worked out, it is possible to suitably substitute $Y_{lm}$ for $E^{ij}$ in 
these formulae.

These equations establish an isomorphism (with all the nice properties like preserving $*$ and $SU(2)$-actions) between the 
convolution algebra $\rho(J)$ ($G^*$) at spin $J$ and the $*$-product algebra of $S_F^2(J)$. That is because we saw that 
$\rho (J)(G^*)$ and $S_F^2(J) \simeq Mat(2J+1)$ are isomorphic, while it is known that $Mat(2J+1)$ and the $*$-product algebra
of $S^2$ at level $J$ are isomorphic.

There are evident generalizations of $P_J$ for other groups and their orbits.

\section{Final Remarks}   

The previous discussions suggest the following coproduct $\delta$ from $Mat(N+1)$ to $Mat(N+1) \otimes Mat(N+1) \, (N = 2J)$:
\begin{eqnarray}
\delta \big( e^{ij}(J) \big) &=& e^{ij}(J) \otimes e^{ij}(J) \,, \nonumber \\
\delta \big( M= M_{ij} e^{ij}(J) \big) &=& M_{ij} \big( e^{ij}(J) \otimes e^{ij}(J) \big) \,, \quad M \in Mat(N+1) \,.
\label{eq:noneqcoproduct}
\end{eqnarray}
$\delta(MN)$ is $\delta(M) \delta(N)$ as we must have and it also preserves $*$ (hermitian conjugation).

But this $\delta$ is not equivariant for any non-trivial group action, such as that of $SU(2)$ or $U(N+1)$, on left, right or by
conjugation. It is different from $\Delta_{KL}$. This example once again illustrates the importance of equivariance for 
discriminating between coproducts.

\vskip 1em
{\bf \large Acknowledgments}
\vskip 1em
We have benefited from inputs from our extended network of colleagues working on fuzzy physics and non-commutative geometry.
We especially thank Denjoe O'Connor, Peter Pre\v{s}najder and Paulo Teotonio-Sobrinho for patient hearings and critical inputs. 
This work was supported by DOE under contract number DE-FG02-85ER40231, by NSF under contract number INT9908763 and by FAPESP, Brasil.

\end{document}